\documentclass[doublecolumn]{aa}
\usepackage{graphicx}

\newcommand{\change}{}

\begin{document}
   \title{Vertical distribution of chromium
    in the atmospheres of HgMn stars
\thanks{Based on observations obtained at the European Southern
Observatory, La Silla and Paranal, Chile (ESO programmes Nos. 62.L-0348, 
65.I-0644, 67.D-0579)}
}

   \author{I. Savanov
          \inst{1,} \inst{2}
          \and
          S. Hubrig\inst{3}}

   \offprints{I.Savanov}

   \institute{Astrophysical Institute Potsdam (AIP),
              An der Sternwarte 16, D-14482 Potsdam, Germany\\
              \email{isavanov@aip.de}
         \and
             Crimean Astrophysical Observatory, Nauchny, Crimea, Ukraine and
             Isaac Newton Institute of Chile, Crimean Branch\\
         \and
             European Southern Observatory, Casilla 19001, Santiago 19, Chile\\
             \email{shubrig@eso.org}
             }

   \date{Received ; accepted }

   \abstract{
We use multiplet 30 \ion{Cr}{ii} lines in the wings of H$_{\beta}$ 
to test the hypothesis of an anomalous concentration of Cr in the upper 
layers of the atmospheres of a sample of 10 HgMn stars. These lines are at
different distances from the H$_\beta$ line center and are therefore formed at 
different depths in the stellar atmosphere.  Except for HD\,49606, all HgMn 
stars show an increase of Cr abundance with height in the stellar atmosphere.
A similar vertical distribution
of Cr, but less pronounced, has been previously found in Am stars.
In contrast, no variation of Cr abundance with the depth has been found for the 
normal late B-type star HD\,196426 and the weak magnetic late B-type star HD\,168733.
It is possible that in HgMn stars the vertical stratification parameter, $a$, depends on 
T$_{\rm eff}$, with the strongest vertical gradient being found in the hotter 
stars.  No correlation was found between $a$ and the average stellar 
abundance $\log \varepsilon$(Cr/H).

   }
\titlerunning{Vertical distribution of chromium in HgMn stars}
   \maketitle
%
\section{Introduction}
The HgMn stars belong to spectral types between A0 and B7 and show
marked peculiarities in atmospheric abundances. The most distinctive
features of their atmospheres are extreme overabundances of Hg (sometimes
exceeding 5~dex) and/or Mn (up to 3 dex). The HgMn stars are slow rotators 
(Abt et al.\ 1972) and appear not to have
strong large-scale organized magnetic fields such as those of other chemically 
peculiar (CP) stars of similar temperature, the Si and He-weak stars.
This does not definitely rule out the presence of more complex fields of
kilogauss order, as suggested by Hubrig and collaborators (Hubrig 
\& Castelli 2001 and references therein).
 
To explain the abundance anomalies in the outer layers of HgMn stars, 
radiatively driven selective diffusion in relatively quiescent atmospheres 
is most often invoked. The HgMn stars are too hot to have a
superficial hydrogen convection zone, and, due to gravitational settling of 
He they are expected not to have He$^+$convective zones. Therefore, 
diffusion occurs in the absence of mixing due 
to convection and of stronger meridional circulations that will occur in faster
rotators.

In contrast to the CP stars with large-scale organized magnetic
fields, the HgMn stars generally do not exhibit periodic variations in 
photometry and spectral line intensities. The interpretation of variability
of magnetic CP stars is that the surface properties (e.g. abundance 
distribution of certain elements, distribution of the magnetic field) are
non-uniform and non-symmetric with respect to their rotation axis.
On the basis of indirect arguments, Hubrig \& Mathys (1995) had suggested that 
at least
two chemical elements (Hg and Mn) might be inhomogeneously distributed over the
surface of HgMn stars. Very recently, for the moderately rotating HgMn star 
$\alpha$\,And periodic line profile changes in the Hg~II $\lambda$ 3984 have
been reported by Wahlgren et al.\ (2001) and Adelman et al.\ (2002), who
could show that the reason for the variability is the non-symmetrical
distribution of Hg over the surface of $\alpha$\,And.

Aside from surface inhomogeneities in the atmospheres of CP stars,
an observational evidence is accumulating in the last years that a number of 
observed spectroscopic features cannot be reproduced with the standard 
model atmosphere codes in which vertical abundance variations are neglected and
where a homogeneous photospheric abundance is assumed (e.g. Ryabchikova et al.\ 2002).
{\change Most observational studies of the abundance stratification have been devoted
to magnetic CP stars and rapidly oscillating Ap stars (e.g. Ryabchikova et al.
2003)}, for which the impact of the 
abundance stratification {\change could be significant} and for which it
has to be taken into account in modelling of pulsational radial velocities,
magnetic fields and for abundance determination.
    
A direct method of determining the vertical stratification is by a comparative
analysis of spectral lines which are formed at different 
depths. Abundances from lines of the same ions formed on either side of
the Balmer jump, in the UV and visual spectral regions (e.g., Alecian 1982; 
Lanz et al.\ 1993), or from the \ion{Cr}{ii} lines of mult.~30 in 
the wings of H$_\beta$  (Savanov et al.\ 2001a), have been determined in a small number 
of CP stars.
The method of abundance analysis using eight \ion{Cr}{ii} lines of the 
same multiplet in the wings of the H$_\beta$ line was introduced by 
Khokhlova \& Topil'skaya (1992).  This presents an excellent opportunity to
investigate the vertical stratification in normal and chemically peculiar stars 
of spectral types from B to F. The \ion{Cr}{ii} lines have
well-determined relative oscillator strengths and cover a fairly
wide wavelength range in the wings of the H$_\beta$ line, being situated at 
distances $\Delta \lambda$ of 1.1\,\AA{} to 49.0\,\AA{} from the line center.
If stratification of Cr is indeed present, and if
this element is overabundant in a thin surface layer, then the dependence of 
the spectral line intensities on $\Delta \lambda$ will be markedly different
from that in a homogeneous atmosphere.
A review of the methods of studying 
vertical stratification in the stellar atmospheres is given in Savanov \&
Kochukhov (1998).

Usually, the abundance anomalies
are ascribed to hydrodynamical processes in the outer stellar layers --
radiatively-driven diffusion, aided or mitigated by magnetic fields,
weak (possibly anisotropic) stellar winds, turbulence, and rotational
mixing.
However, it is difficult to determine the mechanisms 
responsible for abundance anomalies
in the absence of accurate observational information about elemental abundances.
The vertical stratification of Cr has not yet been studied for a representative number
of HgMn stars. Savanov et al.\ (2001a) studied  \ion{Cr}{ii} lines of mult.~30 in the 
wings of the H$_{\beta}$ line in a sample of magnetic CP stars, Am stars 
and one HgMn star (46~Dra). We present in this paper our results of the analysis of abundance 
stratification
in ten slowly rotating HgMn stars. One weak magnetic late B-type star, HD\,168733, and
one normal late B-type star, HD\,196426, have been observed as comparison stars. 
   \begin{table}
      \caption[]{Model atmosphere parameters of the studied stars}
         \label{Mod1}
     $$ 
                  \begin{array}{lllll}
            \hline
            \noalign{\smallskip}
            Star  &  T_{\rm eff} & log g & V sin~i     & Reference \\
            \noalign{\smallskip}
                  &          &       & km s^{-1} &    \\
            \noalign{\smallskip}
            \hline
            \noalign{\smallskip}
            HD~1909^{\mathrm{a}}  & 12400 & 4.00 & 13.0 & Adelman~ et~al.~ 1996        \\
            HD~ 33904 & 12500 & 3.62 & 15.0 & Adelman~ et ~al. ~1996     \\
            HD~ 49606 & 14375 & 3.90 & 15.0 & Adelman~ et~al.~ 1996       \\
            HD~ 71066 & 12100 & 3.95 & 2.0 & Hubrig~ et~ al.~ 1999       \\
            HD~ 78316^{\mathrm{b}} &  13250 & 3.75 & 6.0 & Adelman~ \& Pintado~ 2000 \\
            HD~ 110073 & 12900 & 3.75 & 1.8 & Woolf,~Lambert~ 1999    \\
            HD~ 124740^{\mathrm{c}} & 10350 & 4.00 & 2.0 &  Dolk~ et~ al.~ 2003    \\
            HD~ 165493 & 13890 & 3.90 & 2.8 &  Hubrig~ et~ al.~ 1999    \\
            HD~ 168733 & 13500   & 3.30   & 10.0  &  Lanz~ et ~al.~ 1993        \\
            HD~ 175640 & 12000 & 3.95 & 2.5 &  Hubrig~ et~ al.~1999    \\
            HD~ 178065 & 12200 & 3.54 & 1.5 & Hubrig~ et~ al.~1999    \\
            HD~ 196426 & 12815 & 3.89 & 3.0 & Hubrig~ et~ al.~1999      \\
            \hline
         \end{array}
     $$ 
\begin{list}{}{}{}
\item[$^{\mathrm{a}}$] For HD\,1909B : T$_{\rm eff}$=9000~K, log~g=4.0, V~sin~i= 12~km~s$^{-1}$
(Wahlgren et al.\ 2002)
\item[$^{\mathrm{b}}$] For HD\,78316B : T$_{\rm eff}$=8000~K, log~g=4.0, V~sin~i= 40~km~s$^{-1}$
(Ryabchikova  et al.\ 1998)
\item[$^{\mathrm{c}}$] For HD\,124740B : T$_{\rm eff}$=8000~K, log~g=4.0, V~sin~i= 5~km~s$^{-1}$
(Dolk et al.\ 2003)

\end{list}
\end{table}
%

    
\section{Observations and data reduction}
Spectra of seven HgMn stars in our sample, HD\,33904, HD\,49606, HD\,71066, HD\,78316, 
HD\,110073, HD\,124740 and HD\,165493, were observed on March 14, 1999 at ESO 
with the New Technology Telescope (NTT) and the ESO Multi Mode Instrument (EMMI).
The red arm of EMMI was used in the cross-dispersed echelle mode. The main dispersing
element was the EMMI grating \#14, a 31.6 lines/mm echelle grating mounted in the R4 
configuration. As cross-disperser, the EMMI grism \#5 was employed. The detector was
CCD \#36, a Tektronix with 2048$\times$2048 pixels of 24$\times$24 $\mu$m$^2$.
Setting the entrance slit of the spectrograph to a width of $0\farcs{}8$,
a resolving power $\lambda{}/\Delta{}\lambda{} \approx 7\times10^4$ was achieved over the
whole wavelength range (3970--6620\,\AA).

The data reduction was performed using the ESO image processing package MIDAS. Echelle orders
were automatically detected by a Hough transform, and a two-dimensional
polynomial fit was computed to define their location on the CCD. For both the
scientific and the flat field frames, the background scattered light was modelled by
fitting its level in the interorders by a 2-dimensional cubic spline. The result
was subtracted from the corresponding exposure, prior to the division of the
scientific frame by the flat field. The echelle orders were then extracted from the
resulting frames.

Spectra of the three HgMn  stars HD\,1909, HD\,175640 and HD\,178065, have been
recorded on June 13, 2001 at ESO with the VLT UV-Visual Echelle Spectrograph
UVES at UT2. We used the UVES Dichroic standard settings covering the spectral 
range from 3030\,\AA{} to 10000\,\AA. The slit width was set to $0\farcs{}3$, 
corresponding to a resolving power of $\lambda{}/\Delta{}\lambda{} \approx 1.1\times10^5$.
One weak magnetic late B-type star, HD\,168733, and
one normal late B-type star, HD\,196426, have been observed with the UVES standard setting 
RED~580 covering the spectral 
range from 4800\,\AA{} to 6800\,\AA{} and the entrance slit of the spectrograph set to $0\farcs{}3$
on May 28, 
2000. The spectra have been reduced by the UVES pipeline Data Reduction Software (version 1.4.0),
which is an evolved version of the ECHELLE context of MIDAS.
The signal-to-noise ratios of the resulting NTT and UVES spectra are very high, ranging 
from 200 to 500 per pixel in the one-dimensional spectrum.

    
    \begin{table*}
      \caption[]{The Cr abundances derived from the individual  \ion{Cr}{ii} lines
of mult.~30}
      \label{ResObs}            
     $$ 
        \begin{array}{ccrcccccccccccc}
            \hline
  \lambda &\Delta~\lambda &log~gf  &HD~ &HD~ &HD~ & HD~& HD~&HD  &HD~ &HD~ &HD~ & HD~& HD~ & HD~ \\ 
~\AA &~\AA &(VALD)  &1909  &33904 &49606&71066 &78316& 110073  &124740  &165493 &168733&175640 &178065&196426  \\
            \noalign{\smallskip}
            \hline
            \noalign{\smallskip}
 4812.34 &48.98 &-1.96 &-6.40 &-6.00 &-6.20&-6.30 &-6.30&-5.70 &-5.75 &-6.90 &-5.60&-5.60 &-6.05 &-6.65 \\      
 4824.13 &37.19 &-0.97 &-6.40 &-5.85 &-6.15&-6.10 &-6.30&-5.60 &-5.50 &-6.90 &-5.70&-5.55 &-5.95 &-6.70 \\
 4836.23 &25.09 &-1.96 &-6.20 &-5.90 &-6.20&-6.20 &-6.20&-5.65 &-5.80 &-6.60 &-5.50&-5.55 &-5.95 &-6.70 \\
 4848.24 &13.08 &-1.15 &-6.40 &-5.80 &-6.25&-6.20 &-6.10&-5.65 &-5.10 &-6.60 &-5.60&-5.35 &-5.85 &-6.80 \\
 4856.19 & 5.13 &-2.14 &-6.00 &-5.80 &     &-5.80 &-6.00&-5.40 &-5.50 &      &-5.50&-5.30 &-5.80 &-6.60 \\
 4864.33 & 2.98 &-1.36 &-6.00 &-5.70 &     &-5.75 &-6.00&-5.40 &-5.10 &-6.30 &-5.70&-5.00 &-5.70 &-6.60 \\
 4876.40 &15.08 &-1.46 &-6.30 &-5.80 &     &-6.20 &-5.90&-5.60 &-5.70 &-6.30 &-5.65&-5.40 &-5.95 &-6.75 \\
 4884.61 &23.29 &-2.10 &-6.10 &-5.90 &-6.25&-6.30 &-6.00&-5.70 &-5.80 &-6.90 &-5.55&-5.60 &-6.00 &-6.60 \\
            \noalign{\smallskip}
            \hline
         \end{array}
     $$ 
   \end{table*}
\section{Spectrum synthesis calculations}

Model atmosphere parameters for the studied HgMn stars have already been determined in the past
by us and other authors.
In Table~\ref{Mod1}, for each star,  cols. 2--4 list
effective temperature, surface gravity and the $v\,\sin i$ values.
The last column gives references to the papers in which
the model atmosphere parameters have been determined. Three stars in our sample, HD\,1909, 
HD\,78316
and HD\,124740, are known to be spectroscopic binaries. The adopted parameters of the companions are 
presented in the footnotes of Table~\ref{Mod1}.
Model atmospheres were taken from the Kurucz model
grid {\tt ap00k2odfnew} (Kurucz \cite{K00}).

The spectra were analysed using the spectral synthesis code {\tt STARSP} (Tsymbal 1996).
The details concerning the calculation of synthetical spectra, broadening due to 
the instrumental profile and stellar rotation, and fitting
to the observations can be found in the paper of Savanov et al.\ (2001a).
The Vienna Atomic Line Database {\tt VALD} (Piskunov et al.\ 1995) was used as a primary 
source of the oscillator strengths and other spectral line parameters.
The synthetic spectra of three HgMn stars which are known to be spectroscopic binaries, HD\,1909, 
HD\,78316 and HD\,124740, were calculated and combined using 
the {\tt BINARY} routine in {\tt STARSP}, or with an IDL routine based on 
general relations for composite spectra in binary stars which can be found 
in Savanov et al.\ (2001a).  

Adelman (1994) showed that most HgMn stars have little or no turbulence.
Therefore, in all calculations we assumed zero microturbulent velocity. 
The $v\,\sin i$ values
have been taken either from the referred papers or they were estimated 
independently from the comparison of the observed and computed spectra, after having
degraded the computed spectra for instrumental broadening.
An abundance analysis of \ion{Cr}{ii} was made by matching the observed line profiles 
and computed synthetic spectra.
As all eight \ion{Cr}{ii} lines of the mult. 30 lie in the wings of the hydrogen
H$_{\beta}$ line, particular attention has been paid to drawing the continuum. The adopted
continuum completely relied on the synthetic H$_{\beta}$ profiles, so that
the continuum level for the observed spectra was fixed by the computed 
synthetic spectra in which the wings of the H$_{\beta}$ line profile were regarded as 
local continuum.
To check the correctness of our results, the synthetic spectra have also been calculated with 
two other different codes, {\tt SYNT} (Piskunov 1992) and {\tt SPECTRUM} 
(Gray {\tt http://www1.appstate.edu/}).
The results of the \ion{Cr}{ii} abundances are in good agreement between all three codes. The
difference in synthetic hydrogen line profiles calculated by these codes becomes noticeable   
only in the central part ($\pm$ 2\,\AA{} 
from the center of H$_{\beta}$). However, this part of the line profile is not used in our abundance analysis.

In the previous analyses  of $\beta$\,CrB by Savanov \& Kochukhov (1998) and of 17\,Com by Savanov et al.\
(2001a) the authors studied the
dependence of the accuracy of the Cr abundance
determination on the changes in the stellar parameters and metallicity by running multiple 
syntheses. They could show that the small changes in local continuum due to 
errors in atmospheric parameters or incorporating of additional broadening of the H$_{\beta}$ profile 
do not produce any significant effect on the determinations of parameters 
characterizing the Cr vertical stratification.
The atomic data for the synthesis of \ion{Cr}{ii} lines
of the 30th multiplet used in these previous studies were taken from the VALD database.
The theoretical gf-values given by Kurucz and Bell (1995) supported the relative scale of 
oscillator strengths. In addition, the calculations for the standard stars with no stratification 
of Cr in their atmospheres have been regarded as a check for reliability of the procedure 
of the analysis and of the gf-values.

To estimate the uncertainties in the resulting values of log $\varepsilon$(Cr/H) and the 
accuracy of the fitting of the observed spectrum by the synthetic spectrum,
a grid of synthetical spectra has been calculated for each star assuming a change of $\pm$0.30~dex in 
the Cr abundance.
The values of log $\varepsilon$(Cr/H) for each \ion{Cr}{ii} line have been obtained by quadratic 
interpolation in the grid. As a result, the uncertainty in the fitting procedure introduces
an error of about  0.05\,dex in log $\varepsilon$(Cr/H) for the HgMn stars with the highest Cr 
abundance. The error is by a factor 1.5-2.0 larger for the hot HgMn stars where the \ion{Cr}{ii} lines 
are intrinsically weak or Cr is underabundant.
The results for the Cr abundances of the studied stars derived from different \ion{Cr}{ii} 
lines
of mult.~30 are presented in Table~2. For the two hottest HgMn stars, HD\,49606 and HD\,165493,
a few of the \ion{Cr}{ii} lines which are the closest to the center of the H$_{\beta}$ line are very
weak and could not be used for the abundance analysis.

   \begin{figure*}
   \centering
   \includegraphics[width=17cm]{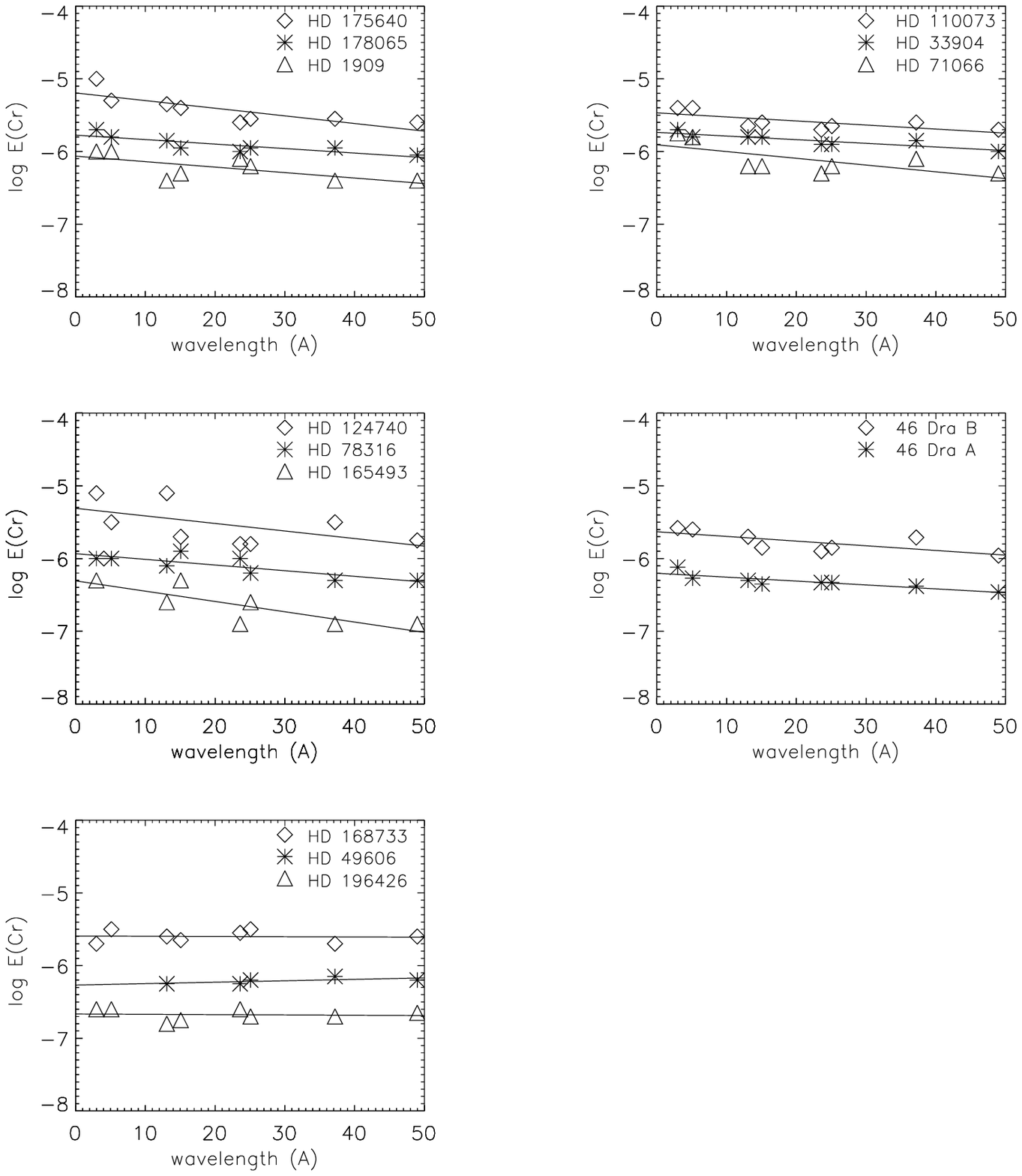}
   \vspace{12mm}
      \caption{Linear approximation of the dependence of the Cr abundance on
               the distance from the center of the H$_{\beta}$ line.
              }
         \label{Fig15}
   \end{figure*}

\section{Results and discussion}

Fig.\ 1 shows plots of the Cr abundances as a function of
distance from H$_\beta$ line center for all 12 stars of our sample including the 
components of the HgMn spectroscopic binary 46~Dra from 
Savanov et al.\ (2001a).  Abundances are shown on a scale where 
$\log \varepsilon$(H)=0.0. 
The results of our analysis are summarized in Table~\ref{Res3}.
In the second column we give the linear regression coefficient $\it{a}$ (the average 
Cr abundance gradient) and its error 
$\sigma_a$ found in an approximation of the Cr abundance as a function of $\Delta{}\lambda$
by the formula {\it log~$\varepsilon$(Cr/H)~= $a$\,$\times$\,$\Delta{}\lambda$~+~b}. 
The coefficient $\it{a}$ is equal to the tangent
of the angle of inclination of the dependence of log $\varepsilon$(Cr/H) on 
$\Delta{}\lambda$ and is a quantitative
characteristic of the vertical Cr abundance gradient. Negative values of $\it{a}$ correspond to
an increase of chromium abundance in upper layers of the stellar atmosphere.
The linear regression coefficient, $\it{a}$, and its error, 
$\sigma_a$, have been determined with an IDL routine which uses the
subroutine SVDFIT. 
This parameter can be regarded as a first approximation to the vertical
distribution of the element in the stellar atmosphere until modeling of
concentration of atoms with optical or geometrical depth can be performed
(e.g. Savanov et al.\ 2001b, Ryabchikova et al.\ 2002).
The last column of Table~\ref{Res3} shows the 'mean' values
of the Cr abundances which were determined from the 
log $\varepsilon$(Cr/H) of the four \ion{Cr}{ii} lines at distances larger than 20\,\AA{} from 
the center of the H$_{\beta}$ line and should be regarded as upper limits for the Cr abundances.

In Table~4 we compare the 'mean' values of the Cr abundance with the results obtained in 
previous studies. The agreement between our abundances and the abundances given in the literature is rather 
good,
although some discrepancies exist for the stars HD\,1909, HD\,49606 and HD\,78316.
The binary nature of HD\,1909 has been recognized by Wahlgren et al.\ (2002). However, the star has been 
considered 
as a single star in the studies of Guthrie (1984) and Adelman et al.\ (1996) who give lower values of Cr 
abundance compared to our determination. For HD\,49606, Adelman et al.\ (1996) determined a higher Cr abundance 
($-$5.89). On the other
hand, our abundance value ($-$6.20) is fully consistent with the determination of Smith \& Dworetsky (1993).
Taking into account the presence of the spectral companion we have derived log $\varepsilon$(Cr/H)=~$-$6.2 
for HD\,78316. This abundance value is lower than the abundance obtained by Adelman \& Pintado (2000),
but higher than that published  by Ryabchikova (1998).
As an additional test of the accuracy of the determination of the 'mean' abundances
we calculated the Cr abundance in the star HD\,175640 using eight unblended \ion{Cr}{ii} lines of 
mult.~44 in the spectral region  4550--4620~\AA.
Calculations were performed with the same source of gf-values (VALD database). We obtained the value of
log $\varepsilon$(Cr/H) is equal to $-$5.55~$\pm$~0.10, very similar to that derived from the 
\ion{Cr}{ii}
lines in the far wings of the H$_{\beta}$ line ($-$5.58).

   \begin{table}
      \caption[]{The average Cr abundance gradient  $\it{a}$, its error $\sigma_a$
      and the derived 'mean' Cr abundance}
      \label{Res3}
     $$ 
         \begin{array}{p{0.5\linewidth}rl} 
            \hline
            \noalign{\smallskip}
            Star  &  (\it{a} \pm \sigma_a)*10^3 & log \varepsilon(Cr/H) \\
            \noalign{\smallskip}
            \hline
            \noalign{\smallskip}
            HD 1909  & -7.5 \pm 3.3 & -6.28 \pm 0.15     \\
            HD 33904 & -5.0 \pm 1.1 & -5.91 \pm 0.06     \\
            HD 49606 & +2.0 \pm 1.4 & -6.20 \pm 0.04     \\
            HD 71066 & -9.3 \pm 4.1 & -6.23 \pm 0.10     \\
            HD 78316 & -7.8 \pm 2.3 & -6.20 \pm 0.14     \\
            HD 110073 & -5.6 \pm 2.2 & -5.66 \pm 0.05     \\
            HD 124740 & -6.2 \pm 1.6 & -5.71 \pm 0.14     \\
            HD 165493 & -14.2 \pm 4.5 & -6.82 \pm 0.15     \\
            HD 168733 & -0.3 \pm 2.1 & -5.59 \pm 0.09     \\
            HD 175640 & -10.5 \pm 3.1 & -5.58 \pm 0.03     \\
            HD 178065 & -6.2  \pm 1.6 & -5.99 \pm 0.05     \\
            HD 196426 & +0.4  \pm 1.9 & -6.66 \pm 0.05     \\
            \hline
            46 Dra A$^{a}$ & -5.4 \pm 1.3 & -6.32 \pm 0.10     \\
            46 Dra B$^{a}$ & -6.4 \pm 2.5 & -5.76 \pm 0.14     \\
            \noalign{\smallskip}
            \hline
         \end{array}
     $$ 
\begin{list}{}{}{}
\item[$^{\mathrm{a}}$] Data from Savanov et al. 2001a.
\end{list}
   \end{table}
%
  \begin{table}
      \caption[]{Cr abundance determinations in the present study and from the literature}
      \label{Res4}
     $$ 
         \begin{array}{p{0.2\linewidth}lll}
            \hline
            \noalign{\smallskip}
            Star   & log \varepsilon & log \varepsilon &  references\\
                    &    (Cr/H)     & (Cr/H)  &\\
                    &    this~paper     &  &\\
            \noalign{\smallskip}
            \hline
            \noalign{\smallskip}
            HD 1909  & -6.28 & -6.40 & Guthrie~ 1984  \\
                     &       & -6.68 & Adelman~ et~ al.~ 1996\\
            HD 33904 & -5.91 &  -5.89 &  Adelman~ \& ~ Pintado~2000\\
            HD 49606 & -6.20 &  -5.89  & Adelman~ et~ al.~ 1996  \\
                     &       & -6.20   & Smith~ \& ~Dworetsky~ 1993  \\
            HD 78316 & -6.20  & -5.89 &  Adelman~ \& ~Pintado~ 2000 \\
                     &        & -6.42 &   Ryabchikova~ 1998\\
            HD 110073 & -5.66  &  -5.50 &  Adelman~ \& ~ Pintado~ 2000\\ 
            HD 175640 &  -5.58 &  -5.50 & Smith~ \& ~Dworetsky~ 1993 \\
            HD 178065 &  -5.99 &  -5.9 &  Guthrie~ 1984\\
                      &        &  -5.84&  Pintado~ \& ~Adelman~ 1996\\
                      &        &  -5.8 &  Zakharova~ 1994 \\  
            HD 196426 &  -6.66 & -6.60  & Smith~ \& ~Dworetsky~ 1993  \\
            \hline
         \end{array}
     $$ 
   \end{table}
The results presented in Fig.~1 and in Table~\ref{Res3} show that except for HD\,49606, 
in all HgMn stars of our sample the Cr abundance in the atmospheres
increases slightly {\change with height in the stellar atmosphere.}
The average Cr abundance gradient $\it{a}$ is
~-0.0076~$\pm$~0.0028. This  corresponds to an abundance increase of 
approximately 0.34 dex in going from the 4812.34~\AA{} line 
to the 4864.33~\AA{} line.

We found no vertical Cr stratification in the atmosphere of the  
HgMn star HD\,49606, which is the hottest HgMn star in our sample. 
The model atmosphere parameters for this star have been taken 
from Adelman et al.\ (1996).  
It has been shown by Smith \& Dworetsky (1993) that the \ion{Cr}{ii} 
lines in the hottest HgMn stars are intrinsically weak. We are not able to 
measure the chromium  abundances from the \ion{Cr}{ii} lines at
4856.19, 4864.33 and 4876.40~\AA, which are the closest to the center of the 
H$_\beta$ line. However, other lines of mult.~30 indicate  a
constant Cr abundance with $\log \varepsilon$(Cr/H)=$-6.20$ (Fig.~1). 
The star HD\,49606 is of special interest because of the observation of a strong longitudinal 
magnetic field ($1.4 \pm 0.2$~kG) reported by Glagolevskij, Panov, \&  
Chunakova\ (1985). However, Hubrig \& Launhardt (1993) could not confirm the 
existence of such a field.
The inspection of old photographic 
spectra revealed splitting of several lines 
into two components probably caused by a companion. The line profiles on our NTT spectrum
show the tendency to be 'square' or rectangular suggesting incipient separation into
two components. Further observations at high resolution and high S/N are needed to decide definitely 
whether HD\,49606 is indeed a spectroscopic binary.

No variations of the Cr abundance with depth have been found for the normal 
late B-type star HD\,196426 and the magnetic star HD\,168733. The longitudinal magnetic field of 
HD\,168733 is rather weak, about $-$600~G, and not detectably variable (Mathys \& Hubrig 1997). 
The lack of stratification of Cr in the magnetic star HD\,168733 is consistent with the results of
the previous analyses of the stars 17\,Com and $\alpha^2$\,CVn (Savanov et al.\ 2001a). 
In contrast, an increase of the Cr abundance
with depth was found for the cool magnetic stars $\beta$\,CrB, $\gamma$\,Equ and 10\,Aql 
(Savanov \&
Kochukhov 1998; Savanov et al.\ 2001a).
The result for $\gamma$\,Equ has been confirmed in the later study by Ryabchikova et al.\ (2002) 
who presented also an evidence for abundance stratification of other 
elements (Ca, Fe, Ba, Si, Na, Pr and Nd).
   \begin{figure}
   \centering
      \includegraphics[width=8cm]{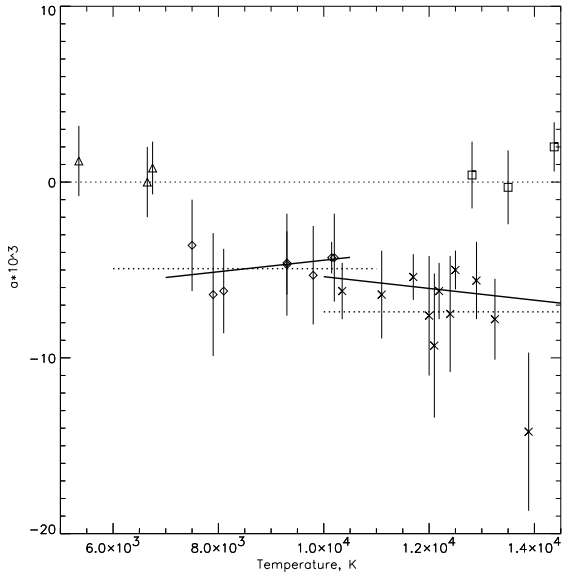}
      \caption{Dependence of the stratification parameter $\it{a}$ on
               T$_{\rm eff}$. The three points (triangles) in the 'cool' end of the diagram are for the 
               standard stars Procyon and $\iota$\,Peg~A and B. Diamonds denote results for Am stars and
               crosses are results from the current investigation of HgMn stars.
               The three open squares correspond to the stars for which no vertical stratification of the 
Cr abundance has been found.
               Dotted lines are mean values for the Am and HgMn groups, {\change while 
two solid lines
               are the regressions for each groups.}
              }
         \label{FigVibStab}
   \end{figure}
%

   \begin{figure}
   \centering

      \includegraphics[width=8cm]{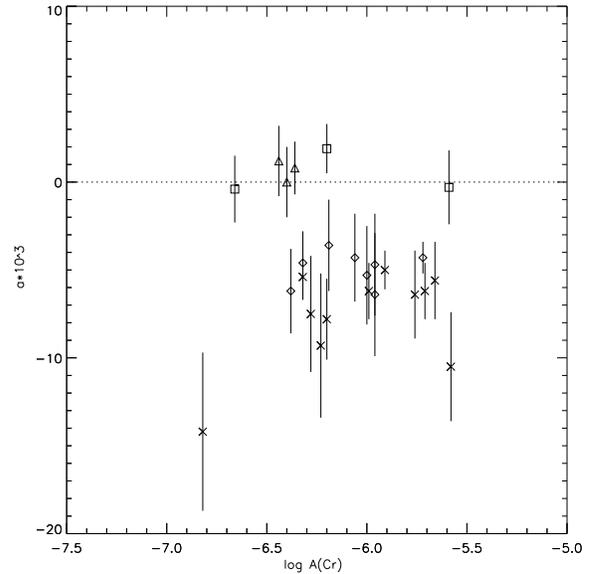}
      \caption{Stratification parameter $\it{a}$ as a function
               of the Cr abundance. Symbols are the same as on Fig. 2.
              }
         \label{FigVibStab2}
   \end{figure}
%


To search for possible general relations, we studied the dependence between 
the vertical stratification parameter $a$ and T$_{\rm eff}$, and between $a$ and 
the Cr abundance $\log \varepsilon$(Cr/H) {\change (Fig. 2, 3).}
In previous studies Procyon and $\iota$\,Peg ~A,~B have been used as standards 
with normal atmospheres and known abundances. As $\iota$\,Peg is a spectroscopic binary,
spectra of the components were calculated and
combined using the BINARY subprogram. 
As expected, no stratification of Cr was found in their atmospheres
(Savanov \& Kochukhov, 1998 and Savanov et al.\ 2001a).
These stars are presented by the three triangles in the 'cool' end of Fig.~2.
The diamonds show the results 
for Am stars which occupy the temperature region from about 7500\,K to 10000\,K. The crosses 
display the results for the HgMn stars, whereas three open squares correspond to 
the stars for which no vertical stratification of Cr has been found (HD\,49606, HD\,196426 and 
HD\,168733). 

There is no significant difference in
$a$ between Am and HgMn stars in the 
temperature region of about 10\,000~K.
From Fig.~2 we may conclude that both groups, Am and HgMn 
stars, show an increase of the Cr abundance in the upper atmospheric layers.
This effect is more pronounced in stars of the HgMn group, for which a mean 
value of $a = (-7.38 \pm 0.78)\times10^{-3} $ was obtained.  For Am stars we find
$a = (-4.93 \pm 0.34)\times10^{-3}$. The mean values for both groups are shown in
Fig.~2 by dotted lines. {\change Two solid lines show the regressions for stars of 
each group. 
It is quite possible that the vertical
stratification parameter $a$ depends on T$_{\rm eff}$ in HgMn stars.} To confirm this 
trend, more observations of HgMn stars on the hotter end are needed. 

In Fig.~3 we plot $\it{a}$ as a function of log$\varepsilon$(Cr/H). 
It appears that the amount of Cr 
stratification ($a \approx -7\times10^{-3}$) does not depend on the average 
abundance, $\log \varepsilon$(Cr/H). One star, HD\,165493, deviates
strongly from this general relation. It is interesting that this hot 
star shows the lowest Cr abundance in the far wings of the H$_\beta$ line 
and the strongest vertical Cr abundance gradient ($14.2 \pm 4.5$)$\times10^{-3}$.
As has been shown by Smith \& Dworetsky (1993) 
the majority of HgMn stars with the T$_{\rm eff} >$ 13000\,K exhibit deficiency of Cr.
The most remarkable example is the star HD\,186122 which is the most Cr deficient star known.
HD\,186122 has been observed with UVES in May 2000,
but the extreme faintness of the \ion{Cr}{ii} 
lines did not allow to carry out the study of Cr abundances
in the wings of the H$_{\beta}$ line. It is possible that the radiative acceleration on
\ion{Cr}{ii} in hotter stars is sufficient to drive it out of the atmosphere. 
{\change Recently, LTE and NLTE radiative accelerations have been calculated for 
different elements including iron--peak elements for a stellar model of 12000\,K 
by Hui--Bon--Hoa et al. (2002). They find that the 
abundances of the iron--peak elements that can be supported by radiation in the 
atmosphere are consistent with the average abundances of HgMn stars
estimated from abundance analyses of individual stars. Additional radiative
force calculations would be useful to clarify the origin of
the underabundance of Cr in the the hotter (T$_{\rm eff} \geq$ 13000\,K) HgMn stars.}

\section{Conclusions}

In comparison to previous studies, our data have been acquired at much higher 
spectral resolution (R$\ge$ 70000)
and higher signal-to-noise ratio (S/N$\ge$ 200). 
Such high quality data allowed us to perform an accurate study even for hotter examples
of HgMn stars in which the \ion{Cr}{ii} lines are intrinsically weak.
The results of our analysis have been compared with the previous work on Am stars and magnetic 
CP stars.

We found indications that there is a vertical chromium distribution
in the atmospheres of nine HgMn stars. 
Since our sample of HgMn stars
includes stars in a 
broad range of effective temperatures with different values of 
surface gravities and different
Cr abundance, we took advantage of this fact to search for possible 
correlations between the parameters describing the vertical stratification and other stellar 
paramaters.
The amount of Cr stratification in 
HgMn stars is similar to that in Am stars. 
Both groups, Am and HgMn stars, 
show an increase of chromium abundance in the upper layers of the 
atmosphere, but this effect is more pronounced for the stars of the HgMn group.
{\change This fact seems to support the hypothesis of a possible relationship between these 
two types of CP stars. On the other hand, the stars of these groups have in some respects
very different properties. E.g. microturbulence is high in Am stars, while it is  
essentially zero in the HgMn stars (Landstreet 1998). Further studies of the Cr abundance
in both types of stars are needed. In particular, it would be very enlightening to 
analyse in the future studies the Cr abundance in the spectral region before and after the 
Balmer jump.}

As steeper slopes are found in hotter HgMn stars, it is quite possible that the vertical
stratification parameter $a$ depends on T$_{\rm eff}$. However, more observations of HgMn 
stars on the hotter 
end are needed to confirm this trend.

{\change We found no vertical Cr stratification 
in the atmospheres of the normal late B--type star HD\,196426, the weak magnetic 
late B--type star HD\,168733 and the hot HgMn star HD\,49606.} It is likely that 
the hot HgMn star HD\,49606 is a
double-lined spectroscopis binary system and the companion should be taken into account in the 
abundance analysis.

\begin{acknowledgements}
We would like to thank F. Castelli for constructive
comments. Part of this work was supported by the German
      \emph{Deut\-sche For\-schungs\-ge\-mein\-schaft, DFG\/} project
      number STR645/1-1. 
\end{acknowledgements}

\end{document}